\documentclass[aps, superscriptaddress, reprint, pre, showkeys,notitlepage]{revtex4-1}
\usepackage{graphicx}
\usepackage{epstopdf}
\usepackage{listings}
\usepackage{xcolor}
\usepackage{amsmath}
\usepackage{amssymb}
\usepackage[caption=false]{subfig}
\usepackage{verbatim}

\begin{document}

\title{Machine learning applied to proton radiography}

\author{Nicholas F. Y. Chen}
\affiliation{
 Clarendon Laboratory, Department of Physics, University of Oxford, Oxford OX1 3PU, United Kingdom
}

\author{Muhammad Firmansyah Kasim}
\affiliation{
 John Adams Institute, Denys Wilkinson Building, Keble Road, Oxford OX1 3RH, United Kingdom
}
\author{Luke Ceurvorst}
\author{Naren Ratan}
\author{James Sadler}
\author{Matthew C. Levy}
\affiliation{
 Clarendon Laboratory, Department of Physics, University of Oxford, Oxford OX1 3PU, United Kingdom
}
\author{Raoul Trines}
\author{Robert Bingham}
\affiliation{
 STFC Rutherford Appleton Laboratory, Chilton, Didcot OX11 0QX, United Kingdom
}

\author{Peter Norreys}
\affiliation{
 Clarendon Laboratory, Department of Physics, University of Oxford, Oxford OX1 3PU, United Kingdom
}
\affiliation{
 STFC Rutherford Appleton Laboratory, Chilton, Didcot OX11 0QX, United Kingdom
}

\date{\today}

\begin{abstract} 
Proton radiography is a technique extensively used to resolve magnetic field structures in high energy density plasmas, revealing a whole variety of interesting phenomena such as magnetic reconnection and collisionless shocks found in astrophysical systems. Existing methods of analyzing proton radiographs give mostly qualitative results or specific quantitative parameters such as magnetic field strength, and recent work showed that the line-integrated transverse magnetic field can be reconstructed in specific regimes where many simplifying assumptions were needed. Using artificial neural networks, we suggest a novel 3-D reconstruction method that works for a more general case. A proof of concept is presented here, with mean reconstruction errors of less than 5 percent even after introducing noise. We demonstrate that over the long term, this approach is more computationally efficient compared to other techniques. We also highlight the need for proton tomography because (i) certain field structures cannot be reconstructed from a single radiograph and (ii) errors can be further reduced when reconstruction is performed on radiographs generated by proton beams fired in different directions.
\end{abstract}


\maketitle

\section{Introduction}

Magnetic fields generated in laser-matter interactions are of primary interest in high energy density physics \cite{laserplasma}. For example, magnetic fields generated by the Weibel instability can explain the collisionless shocks that are found in young galaxies and other astrophysical systems \cite{weibel, astro}. In inertial confinement fusion, magnetic fields are used in one approach to reduce heat losses and thus improve performance of implosions \cite{magicf}, and in another approach (using cylindrical implosions) as a necessary criterion to reach ignition \cite{icfcylindrical}. Also, magnetic reconnection is a commonly studied process  which converts some of the magnetic energy of a system into heat, and understanding the heating mechanism well could lead to better hohlraum design for inertial confinement fusion \cite{magneticreconnection}.

Proton radiography is an extensively used technique that characterizes electric and magnetic fields in plasmas over a wide range of field strengths \cite{roth}. A polyenergetic proton beam, with typical energies on the order of 10 MeV, is usually produced by high intensity laser interaction with solid targets \cite{hedfrontiers}. This beam then interacts with an object of interest (such as plasmas or shock-compressed matter) and gets deflected as a result of the Lorentz force or collisions with atoms \cite{radiographyscatter}. The outgoing beam is captured on a radiochromic film (RCF) stack which can resolve both spatial and energy profiles \cite{rcf}. 

Various methods have been developed in analyzing proton radiographs. Using the principles of differential scattering and stopping, density profiles of dense matter can be retrieved from radiographs \cite{laserdriven}. Via scaling laws, field strengths of electric and magnetic fields  can be estimated \cite{measuringeb,quantitative1,quantitative2}. Also, radiographs can be used to qualitatively understand electric and magnetic field structures \cite{qualitative2,qualitative3,qualitative4}. Furthermore, radiographs can be simulated numerically in order to identify features found in experimental radiographs \cite{levy,compare}.

It is only recently that techniques have been developed to reconstruct fields. The relations between the field structures and proton radiographs have been established by Kugland $\textit{et al}.$ \cite{Kugland} under certain simplifying assumptions, allowing one to obtain the line-integrated transverse magnetic field from a radiograph by solving a 2-D Poisson equation. Graziani $\textit{et al.}$ \cite{morphology} and Kasim $\textit{et al.}$ \cite{kasim} provided extensions to this technique, under similar assumptions. As such, radiographs of systems which do not obey any of the assumptions in \cite{Kugland,morphology,kasim} can only be analyzed qualitatively. 

Machine learning, a field of study which enables the performance of a computer (with respect to a certain task) to increase with its experience, has seen many applications in artificial intelligence problems such as image recognition, recommender systems and speech-to-text \cite{LeCun}. Due to its ability to discover structures in high dimensional data, artificial neural networks (one example of machine learning) has seen many applications in physics, such as analyzing particle accelerator data \cite{particle}, reconstructing images in optical tomography \cite{Kamilov:15} and retrieving 3-D potentials in electron scattering \cite{dynamical}. The flexible nature of artificial neural networks and the prevalence of its usage in image recognition problems prompt us to posit its usage in imaging 3-D magnetic field structures without a need for simplifying assumptions, addressing the gaps found in existing radiograph inversion techniques.

In this paper, we first review existing work on inverting proton radiographs. We then introduce key ideas of artificial neural networks and review their applications in physics. Next, we outline the new method of using artificial neural networks to reconstruct magnetic fields and retrieving field parameters such as characteristic lengths. Via simulations, we show a proof of concept for the above ideas, and discuss how noise and selection of training data affect our results. Using an example, we highlight the need for proton tomography. Finally, we compare the artificial neural network technique with the existing methods of radiograph inversion and suggest a variety of extensions to our research.

\section{Theory}

\subsection{Existing methods of retrieving magnetic fields from radiographs}
\label{subsec:existingmethod}

In this subsection, we will outline the foundational work on proton radiograph inversion by Kugland $\textit{et al.}$, move on to discuss Graziani $\textit{et al.}$ and Kasim $\textit{et al.}$'s extensions, and conclude with the gaps in these methods.

\begin{figure}
\includegraphics[width=0.5\textwidth]{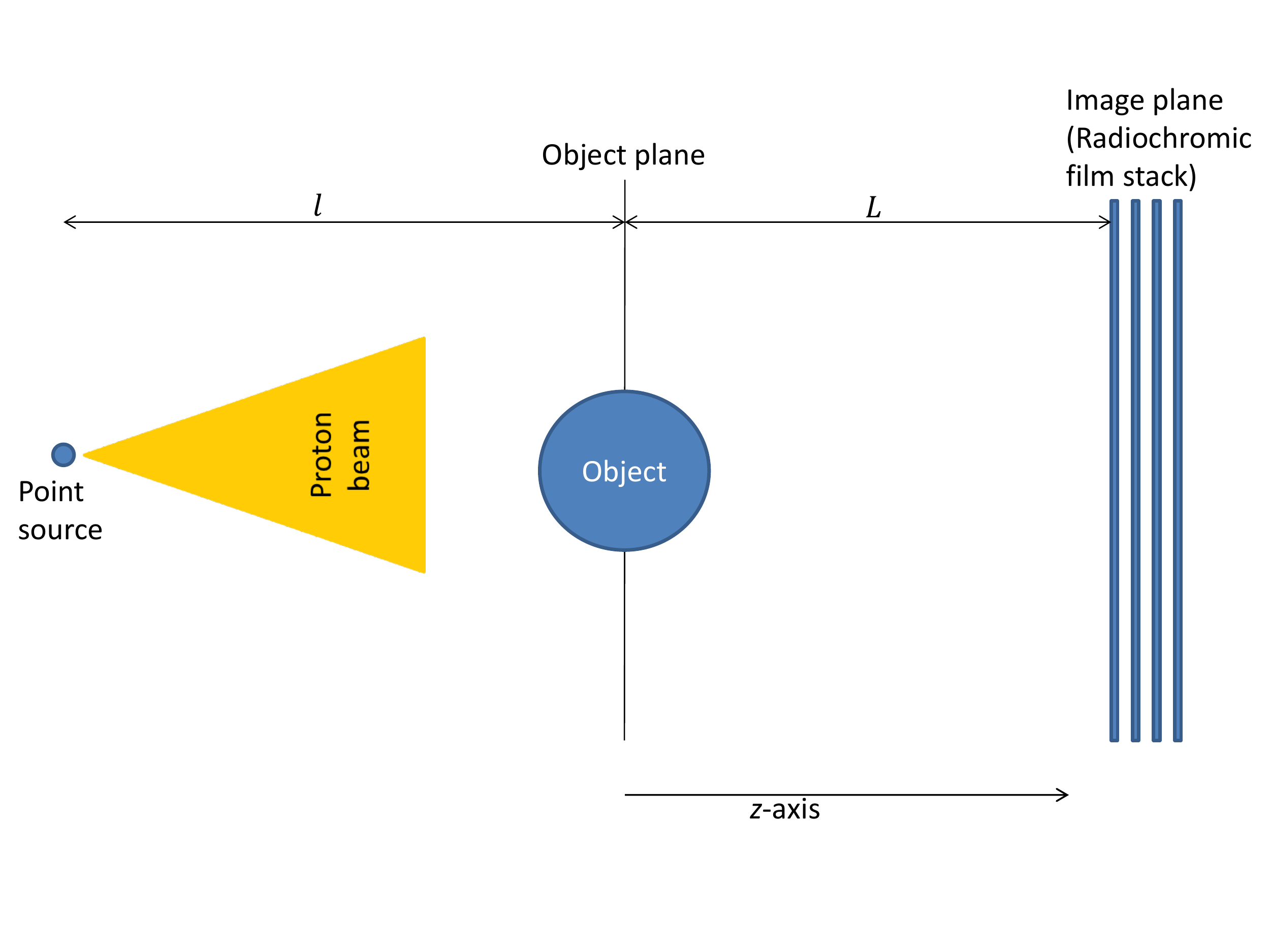}
\caption{Diagram of a typical proton radiography setup. A point source of distance $l$ away from the object emits a beam of protons moving generally in the $z$-direction. $L$ is the distance between the object and image plane.}
\label{fig:setup}
\end{figure}

First, we go through Kugland $\textit{et al.}$'s \cite{Kugland} definitions: The coordinates are defined such that the object is placed at $z=0$ (object plane), and $(x, y)$ refers to the coordinates on the image plane (see Fig. \ref{fig:setup}). At the object plane, the proton's coordinates are denoted as $(x_0,y_0)$. The distance between the proton source and the object is $l$ while the distance between the object to the image plane (radiochromic film stack) is $L$. $a$, the characteristic length of the object, is assumed to be much smaller than $l$ (paraxial limit) and $L\gg l$ for high magnification.

In order to get a tractable result, Kugland $\textit{et al.}$ have made some simplifying assumptions. We start off with those relating to the proton source: (i) The source can be treated as a point source. Else, the radiograph will be blurred and the resolution of field structures will be affected. (ii) The protons deviate from their straight-line trajectories solely due to the Lorentz force interaction with the object, and we can ignore space-charge effects because the beam is charge-neutral as a result of co-moving electrons \cite{comovingelectrons}. (iii) The angular width of the beam is much greater than $a/l$ so that intensity variations in the image plane are due to proton interactions with the object, and not the angular distribution of the proton beam. 

Consider the dimensionless parameter

\begin{eqnarray}
\mu\equiv \frac{l\beta}{a},
\end{eqnarray}

\noindent where $a$ is a characteristic length of the electromagnetic field, and $\beta$ is a characteristic deflection angle. One core assumption in Kugland $\textit{et al.}$ is that $\mu\ll1$ (hence known as the linear regime), where the spatial variation of the intensity on the screen is small. This is in contrast to the non-linear regime ($\mu$ on the order of 1 or more) where the intensity variations are large, leading to non-linear features. One example of non-linear features is caustics, which occurs when the Jacobian determinant

\begin{eqnarray}
 \left|\frac{\partial (x,y)}{\partial (x_0,y_0)}\right|=0,
\end{eqnarray}

\noindent resulting in features of high intensity (usually multiples of the background intensity).

Furthermore, assuming that the velocity of the proton $\mathbf{v}$ is approximately constant while the proton is within the object, (trajectories are not perturbed within the plasma so $\mathrm{dt=d}z_0/v$), the only relevant component of the magnetic potential is the one in the $z$-direction, $A_z$. Defining the line-integrated potential as

\begin{eqnarray}
\Phi(x_0,y_0)= \int_{-\infty}^{\infty}A_z(x_0,y_0,z_0)\mathrm{d}z_0,
\end{eqnarray}

\noindent then with all the assumptions listed above, Kugland $et\:al.$'s formula for radiograph inversion reads:

\begin{eqnarray}
 \mathbf{\nabla_\perp^2}\Phi(x_0,y_0) = \frac{\sqrt{2m_p K}}{el} \Big( 1-\frac{I}{I_0}\frac{L^2}{l^2}\Big),
\label{eqn:poisson}
\end{eqnarray}

\noindent where $\nabla_\perp$ is the gradient with respect to the transverse coordinates $(x_0,y_0)$, $m_p$ is the mass of the proton, $K$ is the (non-relativistic) kinetic energy of the proton, $e$ is the charge of an electron, $I$ is the proton intensity distribution at the image plane and $I_0$ is the proton intensity distribution in the object plane. As such, given the intensity profile at the object plane $I_0(x_0, y_0)$ and radiograph intensity profile $I(x, y)$ (which can be transformed to $I(x_0,y_0)$ via the mapping $x=\frac{L}{l}x_0,y=\frac{L}{l}y_0$) in the regime $\mu\ll 1$, one can solve a 2-D Poisson equation to get the line-integrated potential $\Phi(x_0,y_0)$, thereby allowing one to reconstruct the line-integrated transverse magnetic field.

Using a series of perturbations, and assuming the linear regime $\mu \ll 1$, Graziani $\textit{et al.}$ \cite{morphology} proposed a correction term in the right hand side of equation (20) in Kugland $\textit{et al.}$ (equation (\ref{eqn:poisson}) in this paper) which leads to a second-order non-linear partial differential equation. The authors then conducted a simulation of their proposed equation, and found that their method reconstructed the line-integrated magnetic field accurately at locations near the peak field strength, but was inaccurate at locations where the field strengths are at least 3 orders of magnitude less than the peak field strength. Also, Graziani $\textit{et al.}$ briefly sketched a method to retrieve the line-integrated magnetic field in the non-linear regime, assuming that the direction of the proton trajectory within the object is nearly constant. 

Another method, based on computational geometry, was implemented by Kasim $\textit{et al.}$ \cite{kasim}. This method works well in the beginning of the caustic regime (early part of the non-linear regime where $\mu\textgreater 1$), but the relative errors start to become very large in the regime of branching caustics (later part of the non-linear regime). Also,  Kasim $\textit{et al.}$ demonstrated the large errors that come with solving the Poisson equation in Kugland $\textit{et al.}$ for a system in the non-linear regime. 

So far, we have seen that existing methods of magnetic field reconstruction require simplifying assumptions in order to get an equation which, when solved, gives the line-integrated transverse magnetic field. This highlights two gaps: (i) In later parts of the non-linear regime (e.g. branching caustics regime), there is no known reconstruction method despite the fact that non-linear features do occur in some experimental radiographs \cite{nonlinear1,nonlinear2,nonlinear3}. In this regime, experimental radiographs are analyzed by comparison to simulated radiographs of a hypothesized magnetic field structure. (ii) In both regimes, there is no reconstruction method that can give the 3-D magnetic field. As we will demonstrate in the next few subsections, the proposed artificial neural network method can address both gaps.

\subsection{Artificial neural networks (ANN)}

\begin{figure}
\includegraphics[width=0.5\textwidth]{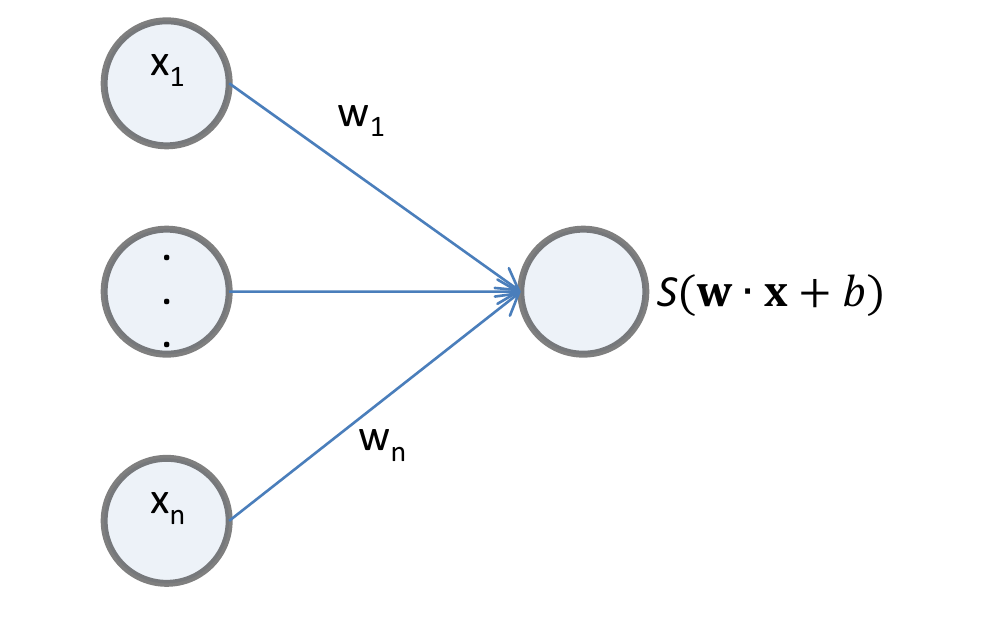}
\caption{A neuron, the basic unit in an artificial neural network. The input vector $\mathbf{x}$ is mapped to a scalar $y$ via a non-linear function $S$. The connections represent the inputting of elements of $\mathbf{x}$ into the neuron, and each connection is assigned a weight, which is used in calculating the output.}
\label{fig:neuron}
\end{figure}

Artificial neural networks are a class of models inspired by biological neural networks, commonly used for tasks that are too complicated for rule-based programming.  A neuron, the basic unit of an artificial neural network, takes an input vector $\mathbf{x}$, and returns a scalar output $y$ given by

\begin{eqnarray}
y=S(\mathbf{w}\cdot \mathbf{x} + b),
\end{eqnarray}

\noindent where $\mathbf{w}$ is the weight vector of the neuron, $b$ is a bias term and $S$ is an activation (or transfer) function, usually a non-linear function such as the sigmoid function $s(t)=\frac{1}{1+e^{-t}}$ or the hyperbolic tangent (see Fig. \ref{fig:neuron}). 

\begin{figure}
\includegraphics[width=0.5\textwidth]{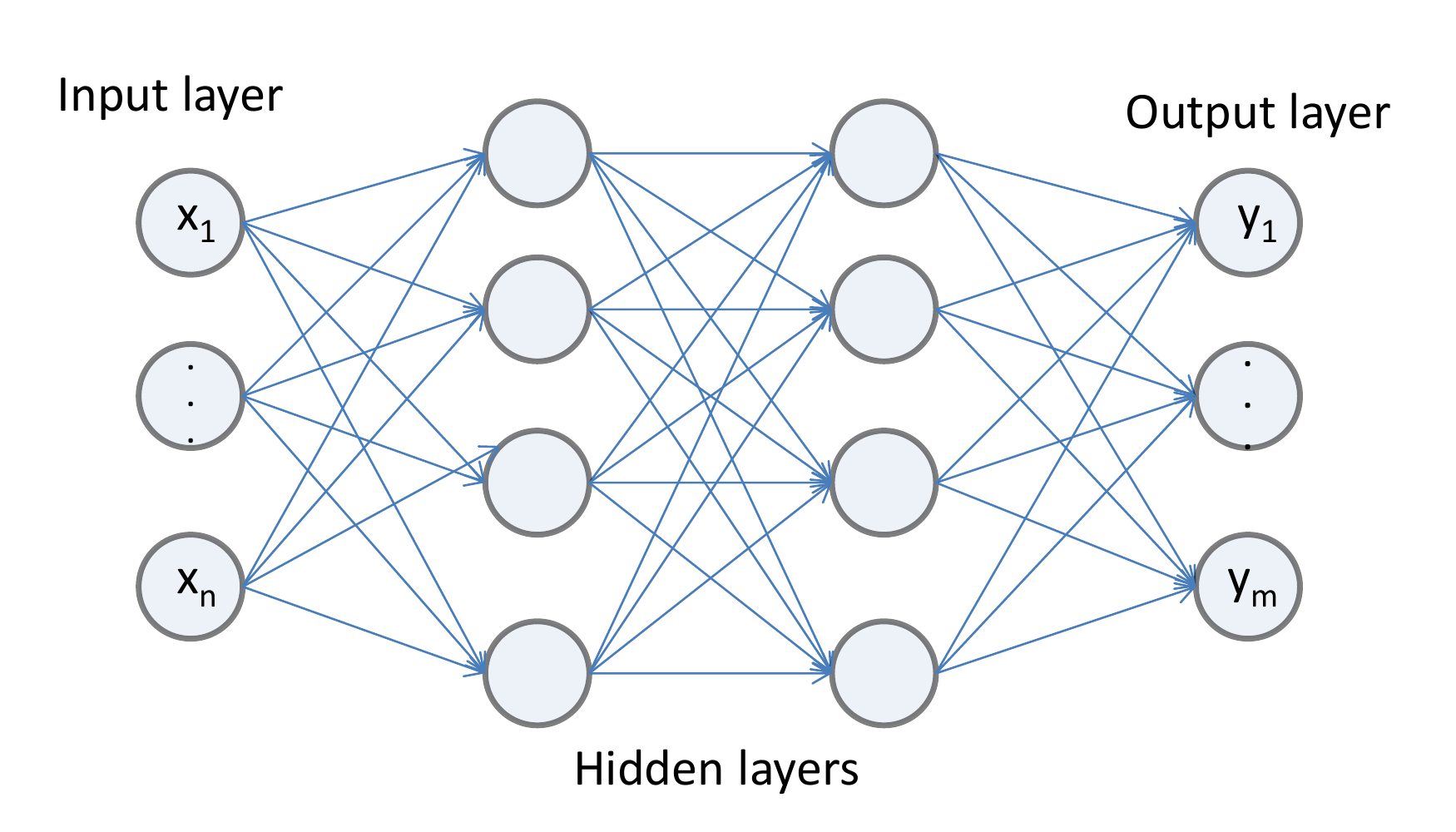}
\caption{Schematic of a feedforward neural network, where the outputs of one layer are fed into the inputs of an adjacent layer. It takes a vector $\mathbf{x}\in\mathbb{R}^n$ and outputs a vector $\mathbf{y(x,\theta)}\in\mathbb{R}^m$, where $\mathbf{\theta}$ represents the parameters (weights and biases) of the neural network.}
\label{fig:nn}
\end{figure}

An artificial neural network consists of many of these neurons joined by the inputs and outputs (i.e. the output of several neurons is fed into the input of another neuron), and parameterized by $\mathbf{\theta}$, the vector of weights and biases. There are many variants of artificial neural networks such as recurrent neural networks, cascade-forward networks, and the feedforward neural network. A typical feedforward neural network consists of an input layer with $n$ inputs, hidden layers of arbitrary numbers of neurons and an output layer of $m$ neurons (see Fig. \ref{fig:nn}). Information only moves forward (hence the name feedforward), and between adjacent layers: The output from the $L^{th}$ layer strictly goes into the input of the $(L+1)^{th}$ layer. Essentially, the feedforward neural network is a function that maps a vector in $\mathbb{R}^n$ to a vector in $\mathbb{R}^m$. 

One very powerful feature of feedforward artificial neural networks is that with just a single hidden layer, it can approximate any Borel measurable function (which includes any continuous function) to any desired degree of accuracy given enough neurons in the hidden layer \cite{hornik}. Furthermore, any real-valued continuous function can be approximated arbitrarily well if we use (i) any continuous non-constant activation function or (ii) a (not necessarily continuous) squashing function as an activation function. A squashing function $\Psi$ is defined as one that is non-decreasing, $\lim_{\lambda \to \infty}\Psi (\lambda)=1$ and $\lim_{\lambda \to -\infty}\Psi (\lambda)=0$. As such, feedforward artificial neural networks are universal approximators so if we encounter errors during application, it must be due to (i) inadequate learning, which can arise from insufficient training data or improper implementation of the training process (more on the training process in the next paragraph), (ii) insufficient neurons in the hidden layer, or (iii) a stochastic relationship between inputs and the desired outputs.

While \cite{hornik} established that feedforward artificial neural networks can express any Borel measurable function, we also need to know how to select the appropriate $\mathbf{\theta}$ so that the artificial neural network is a good approximation to our desired function $\mathbf{g(x)}$. Given data on inputs $\mathbf{x}$ and targets $\mathbf{g(x)}$, we can train the artificial neural network: Upon specifying an objective function (e.g. mean-squared error between the target $\mathbf{g(x)}$ and the output of the artificial neural network $\mathbf{y(x,\theta)}$), we can use backpropagation \cite{backprop} (a method of finding the gradient of the objective function with respect to $\mathbf{\theta}$) in conjunction with an optimization algorithm (e.g. stochastic gradient descent) to iteratively adjust $\mathbf{\theta}$ until the objective function is minimized (either locally or globally). 

Considering that $\mathbf{\theta}$ is high-dimensional, optimization can be difficult. However, Choromanska $\textit{et al.}$ \cite{losssurface} transformed common objective functions to a physical problem explored in \cite{auffinger1,auffinger2} and concluded that for large size networks, most local minima are equivalent and the resulting artificial neural networks approximate the target function with similar accuracies. Furthermore, there is not much value in finding the global minimum to the objective function because it could lead to overfitting (modeling noise in the data).

Given the merits of artificial neural networks mentioned above, it is no surprise that artificial neural networks are being used in some areas of physics. In astrophysics, artificial neural networks are being used to separate astrophysical signals from the cosmic microwave background \cite{astrosignal} and to classify stars and galaxies \cite{astrogalaxy}. In optical tomography, Kamilov $\textit{et al.}$ \cite{Kamilov:15} used artificial neural networks to enhance the beam propagation method \cite{VanRoey:81} and retrieve the index of refraction of a 3-D object. The images reconstructed from this technique were found to be of higher quality than images from optical diffraction tomography and Radon tomographic reconstruction. In dynamical electron scattering experiments, Van den Broek $et\:al.$ used artificial neural networks to retrieve 3-D potentials \cite{dynamical}.

In terms of retrieving information from proton radiographs, artificial neural networks can be used as an input/output map. Given the radiograph pixel values as inputs, the artificial neural network should output useful quantities related to the $\mathbf{B}$ field. Such an artificial neural network can be trained (i.e. the weights are selected) by applying backpropagation and optimization algorithms to many sets of input (radiograph pixels) and target ($\mathbf{B}$ field quantities) examples. After enough training examples, the artificial neural network will become a tool that can estimate $\mathbf{B}$ field quantities given a radiograph. One key strength of the artificial neural network method is that the bulk of the computational cost comes from generating training data and training the artificial neural network, which is a one-off cost.

\section{Methods}

\subsection{Reconstruction of an arbitrary $\mathbf{B}$ field}
\label{subsec:recon}

\begin{figure*}
\includegraphics[width=1\textwidth]{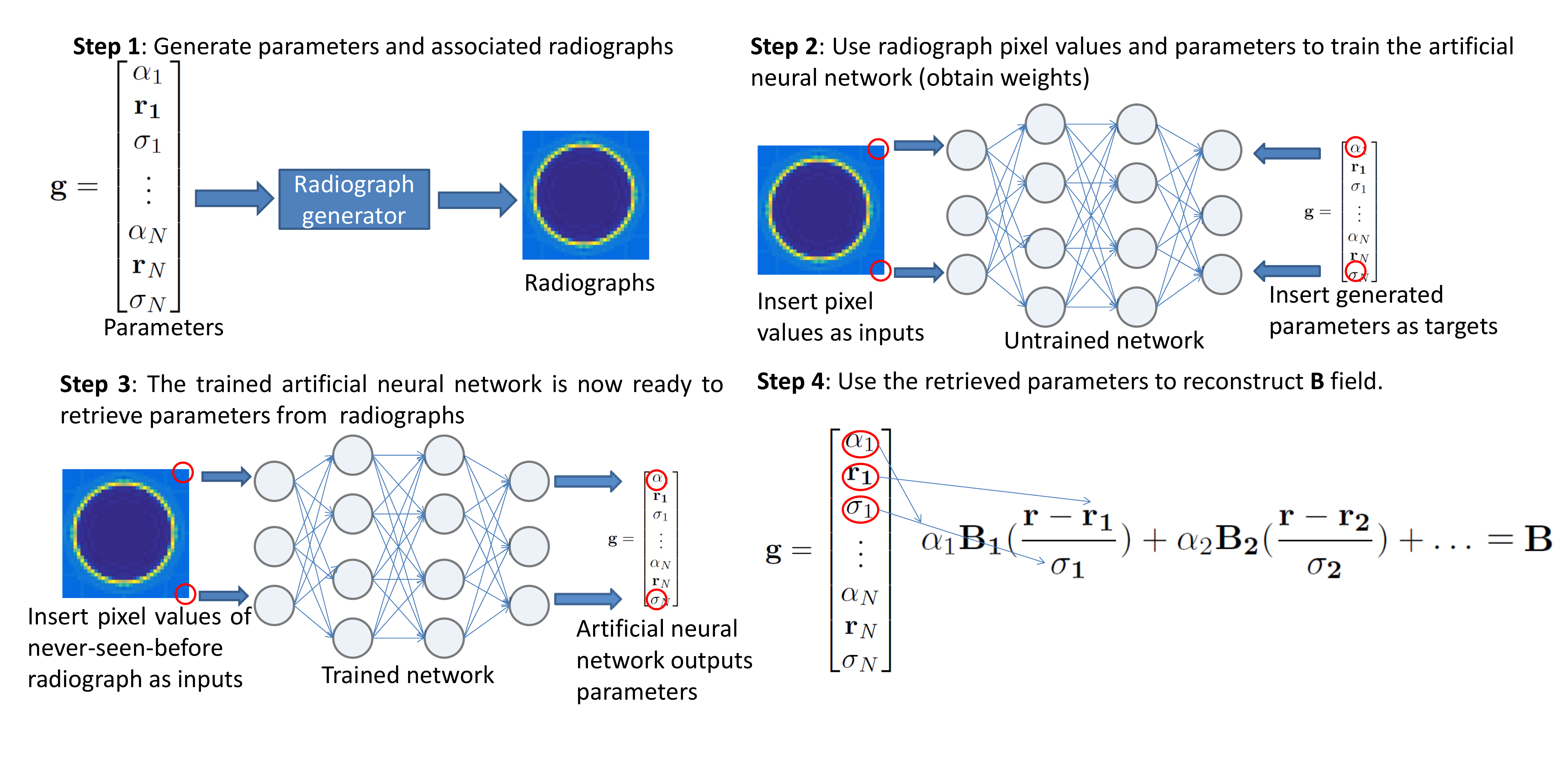}
\caption{Schematic of the prediction and reconstruction process.}
\label{fig:predictreconstructprocess}
\end{figure*}

In this section, we outline the steps to using an artificial neural network to reconstruct any arbitrary $\mathbf{B}$ field.  First, we expand the magnetic field \textbf{B(r)} as a linear combination

\begin{eqnarray}
\mathbf{B(r)} =&& \alpha_1 \mathbf{B_1}(\mathbf{\frac{r-r_1}{\sigma_1}}) + \alpha_2 \mathbf{B_2}(\mathbf{\frac{r-r_2}{\sigma_2}})  + \ldots  \\
=&&\sum_{n=1}^{N}  \alpha_n \mathbf{B_\textit{n}}(\mathbf{\frac{r-r_\textit{n}}{\sigma_\textit{n}}}),
\label{eqn:expansion}
\end{eqnarray}

\noindent where $N$ is the number of terms used in the expansion, $\alpha_n$ is a scalar coefficient for the $n^\textrm{th}$ term, $\mathbf{B_\textit{n}}$ is a `basis' magnetic field, $\mathbf{r_\textit{n}}$ is the position offset of the field and $\sigma_\textit{n}$ is a scaling factor. While not necessary, these basis fields should be chosen such that most magnetic fields in plasmas can be represented with as few basis fields as possible, so that we require less training data to train the artificial neural network. One possible way to achieve this is to use principal components analysis (PCA) \cite{pca} on a large dataset of known $\mathbf{B}$ fields in plasmas. Principal components analysis looks at a large set of multidimensional vectors and first finds the direction of highest variance in the data (the first principal component), and then finds a set of vectors orthogonal to the first principal component that explains the remainder of the variance. While $\mathbf{B(r)}$ is a vector field, it can be converted into a vector $\mathbf{c}$ for principal components analysis by concatenating the magnetic fields at various different points, e.g. for a grid that runs from 0-9 in the $x$, $y$ and $z$ directions, we can write

\begin{eqnarray}
\mathbf{c} = 
\begin{bmatrix}
\mathbf{B(r_{000})} \\ \mathbf{B(r_{001})} \\ \vdots \\ \mathbf{B(r_{999})}
\end{bmatrix},
\label{eqn:vectorc}
\end{eqnarray}

\noindent where $\mathbf{r_\textit{xyz}}$ is the vector ($x, y, z$). If such convenient basis fields cannot be determined, we can use the fact that all magnetic fields can be written in the form of equation (\ref{eqn:vectorc}), and let each element of the vector correspond to a basis field (i.e. $\mathbf{B_1}$ corresponds to $\mathbf{B_x(r_{000})}$, $\mathbf{B_2}$ corresponds to $\mathbf{B_y(r_{000})}$ and so on).

Next, generate training data by creating variations of the parameters $\alpha_n, \sigma_n \text{ and } \mathbf{r_\textit{n}}$ in the form

\begin{eqnarray}
\mathbf{g} = 
\begin{bmatrix}
\alpha_1 \\ \mathbf{r_1} \\ \sigma_1 \\ \vdots \\  \alpha_N \\ \mathbf{r_\textit{N}} \\ \sigma_N
\end{bmatrix},
\label{eqn:parameters}
\end{eqnarray}

\noindent and then conducting numerical simulations (e.g. using software packages mentioned in \cite{levy,compare}) to obtain the radiograph for each variation of the parameters. The radiograph is expressed as a vector where each element represents the intensity of the protons at a specific pixel.

Then, using the radiograph pixel values as inputs and $\mathbf{g}$ as the targets, apply backpropagation and optimization algorithms to train the feedforward neural network. After training, the artificial neural network is ready to reconstruct $\mathbf{B}$ fields: Input the radiograph into the artificial neural network to obtain the predicted parameters (in the form of equation (\ref{eqn:parameters})), and insert these values into equation (\ref{eqn:expansion}). See Fig. \ref{fig:predictreconstructprocess} for a schematic of the training and reconstruction process.

\subsection{Assumptions, practical considerations and implementation}

In our simulations, we have made some assumptions for simplicity, but these assumptions are not crucial in the success of our approach. We assumed that the probe beam only interacts with the plasma via the $\mathbf{B}$ field (no electric fields or collisions with matter). We also assumed that the proton source is a point source, and the probe beam is a planar sheet (velocities in the $z$ direction, before deflection from the plasma, are uniform). As feedforward artificial neural networks are universal function approximators, in principle the technique outlined in the previous section will still work even if the assumptions are violated (e.g. protons interact with the electric field of the plasma, protons collide with the plasma, proton source is of finite size, probe beam follows a specific angular distribution), as long as we include these effects during the production of training data (radiographs). 

To obtain the radiographs we start off with the Lorentz force equation for $\mathbf{B}$ fields only, given by

\begin{eqnarray}
\frac{\mathrm{d}\mathbf{v}}{\mathrm{d}t}=\frac{e}{m_p}\mathbf{v}\times\mathbf{B}.
\end{eqnarray}

\noindent This equation, along with $\frac{\mathrm{d}\mathbf{r}}{\mathrm{d}t}=\mathbf{v}$ was numerically integrated given the initial conditions of $\mathbf{r}$ and $\mathbf{v}$ to get the final positions of the protons on the screen. These final coordinates are then binned in order to produce radiographs.

We used a fully connected (dense) feedforward artificial neural network for simplicity, and we will discuss the possibilities of using other types of artificial neural networks in section \ref{sec:future}. Scaled conjugate gradient was the optimization algorithm of choice during training because: (i) it is not RAM intensive (this is an important factor because in order to get more accurate results, training with more complicated $\mathbf{B}$ fields and higher resolution radiographs are required, resulting in an increase in the number of weights in the artificial neural network. If the optimization algorithm does not scale well, an impractical amount of RAM will be required); (ii) it can take advantage of parallel CPU and GPU computing, allowing it to run effectively on supercomputing clusters.

Before the training process, the entire data set is scaled so that each feature of the input and target (e.g. $\sigma_1$, $\alpha_1$, the proton intensity in pixel 1 etc.) falls in the range [-1,1] to prevent features of small magnitude from converging slowly during optimization \cite{efficientbackprop}, and the scaling is undone afterwards. The objective function was chosen to be the mean squared error (MSE) between the artificial neural network output and the target. 

\begin{figure}
\includegraphics[width=0.5\textwidth]{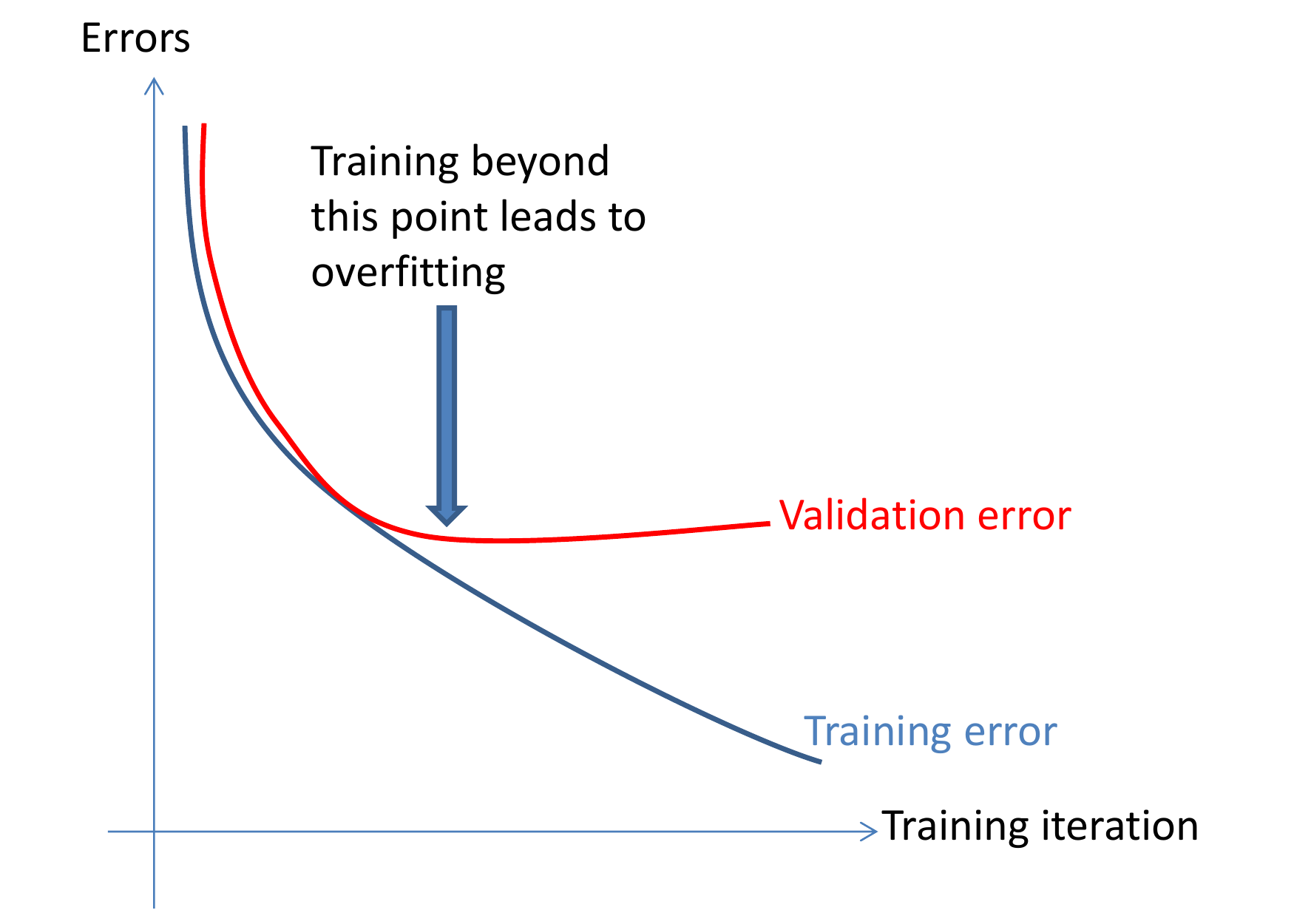}
\caption{Typical curves of training and validation errors with respect to training iteration. Beyond a certain point, the artificial neural network starts to model noise, causing the validation error to increase. Training should be halted when this happens.}
\label{fig:tvt}
\end{figure}

Due to the flexibility of artificial neural networks, overfitting (accidental modeling of noise) is an issue so early stopping and neural network regularization are implemented. In early stopping, training is halted when the errors starts increasing on a data set that was not used in the training process \cite{earlystopping}. This is done by first splitting the entire simulated data set into training, validation and testing sets at random in the ratio 70/15/15. The artificial neural network is applied to the training set, and during each iteration the mean squared error for the validation set is calculated. Initially, after each iteration, the artificial neural network becomes better at modeling the physical phenomenon and the validation mean squared error will decrease. There will come to a point when the artificial neural network starts to model the noise in the training set, and the validation mean squared error will stop decreasing and eventually start increasing (See Fig. \ref{fig:tvt} for an illustration).  Training is halted after a specified number of iterations fail to decrease the validation mean squared error. In neural network regularization \cite{weightdecay}, the objective function is modified by adding a term proportional to the mean squared weight, and the constant of proportionality is known as the regularization parameter (chosen via cross-validation). This penalizes the neural network for having large weights or too many neurons, thus encouraging simpler models. 

At this point, the artificial neural network is used to predict quantities on the testing set, and the testing errors are indicative of the artificial neural network's overall performance. For each simulated data set, the training process is run with the number of neurons in the hidden layer ranging from 10 to 100 in steps of 10, and the configuration with the lowest value of the objective function (mean squared error plus regularization term) is initially picked. If the value of the objective function is still decreasing when 100 neurons are used, then the search is extended in steps of 10 till 150 neurons. Once a single layer configuration is picked, another search is performed with multiple hidden layers (in increments of one layer), up to 5 hidden layers. Similarly, if the value of the objective function is still decreasing when 5 hidden layers are used, the search is extended to 10 hidden layers. After this search, the configuration with the lowest value of the objective function is picked and reported in the results section.

\subsection{Retrieval of specific parameters}
\label{subsec:retrieval}

The idea presented in section \ref{subsec:recon} requires large amounts of data and processing power, and might be more than necessary if the user only intends to retrieve certain parameters of the $\mathbf{B}$ field, such as the peak field strength or the full width half max (FWHM) of a Gaussian magnetic flux rope, instead of reconstructing the entire field. This assumes that the user already knows the remainder of the parameters beforehand. For example, if the user only wants to retrieve the peak $\mathbf{B}$ field strength (proportional to the $\alpha_i$ coefficient), then the model of the $\mathbf{B}$ field, the offset $\mathbf{r_i}$ and the scaling factor $\sigma_i$ must be known. In this case, the user can repeat the procedure in section \ref{subsec:recon}, except with the following changes: (i) data is generated by varying only the parameter(s) of interest; (ii) the target vector consists of only the parameter(s) of interest. In fact, this can be applied to parameters other than $\alpha_i, \mathbf{r_i}$ and $\sigma_i$. For example, in an ellipsoidal magnetic blob (which is a spheroid), there are two characteristic lengths, one characterizing the length in the $xy$ plane $a$ and the other characterizing the length along the $z$-axis $b$. If the user knows all other parameters  and wants to retrieve $a$ and $b$, then an artificial neural network trained on simulated radiographs which variation is only due to varying values of $a$ and $b$ will do the job.

\section{Results and discussion}

We have simulated special cases of equation (\ref{eqn:parameters}) as a proof of concept of the idea in section \ref{subsec:recon}. All results shown here come from applying a trained artificial neural network on the testing set (which is not used in training the artificial neural network), and is indicative of the performance when tested on experimental data. Some of the simulation parameters used in the following subsections can be found in table \ref{table:params}.

\begin{table*}[]
\centering
\begin{tabular}{|c|c|c|}
\hline
\multicolumn{1}{|l|}{} & Value in subsection A-E & Value in subsection F \\ \hline
Ellipsoidal blob parameter $a$/mm & 0.1 & 0.7 \\ \hline
Ellipsoidal blob parameter $b$/mm & \multicolumn{2}{c|}{1} \\ \hline
Flux rope height/mm & 2 & 0.3 \\ \hline
Flux rope parameter $a$/mm & 0.8 & 0.5 \\ \hline
Distance between proton source and object $l$/mm & \multicolumn{2}{c|}{7} \\ \hline
Distance between object and screen $L$/mm & \multicolumn{2}{c|}{93} \\ \hline
Number of neurons in input layer & 2500 & 5000 \\ \hline
Number of neurons in output layer & \multicolumn{2}{c|}{2 for subsections A, B, D, F and 1 for subsections C, E } \\ \hline
Velocity of protons in the $z$ direction/ms$^{-1}$ & \multicolumn{2}{c|}{$10^6$} \\ \hline
Velocity of protons in the $x,\:y$ direction/ms$^{-1}$ & Ranges from -5$\times 10^4$ to 5$\times 10^4$ & Ranges from -6.9$\times 10^5$ to 6.9$\times 10^5$\\ \hline
\end{tabular}
\caption{Parameters for simulations in the following subsections. The proton velocities for subsection F refer to protons in the beam fired in the $z$ direction.}
\label{table:params}
\end{table*}

\subsection{Reconstructing magnetic fields, a proof of concept}

Consider the following two fields: (a) a magnetic ellipsoidal blob, representative of fields generated by the Weibel instability \cite{weibel}, that can be written as

\begin{eqnarray}
B_\phi = B_0\frac{r_0}{a}\mathrm{exp}(-(\frac{r_0^2}{a^2}+\frac{z_0^2}{b^2})),
\label{eqn:blob}
\end{eqnarray}

\noindent where $B_0$ is proportional to the peak field strength, $r_0$ is the distance to the center in the $xy$ plane, $z_0$ is the distance to the center along the $z$-axis, and $a,\:b$ are characteristic lengths of the ellipsoid; (b) a magnetic flux rope of Gaussian cross section, representative of fields due to laser generated plasma flows \cite{fluxrope}, can be written as

\begin{eqnarray}
B_y = B_0\mathrm{exp}(-\frac{x_0^2+z_0^2}{a^2}),
\end{eqnarray}

\noindent where $B_0$ is the peak field strength, $x_0$ and $z_0$ are the distances to the center along the $x$- and $z$- axes respectively, and $a$ is a characteristic length of the Gaussian.

\begin{figure}
\includegraphics[width=0.5\textwidth]{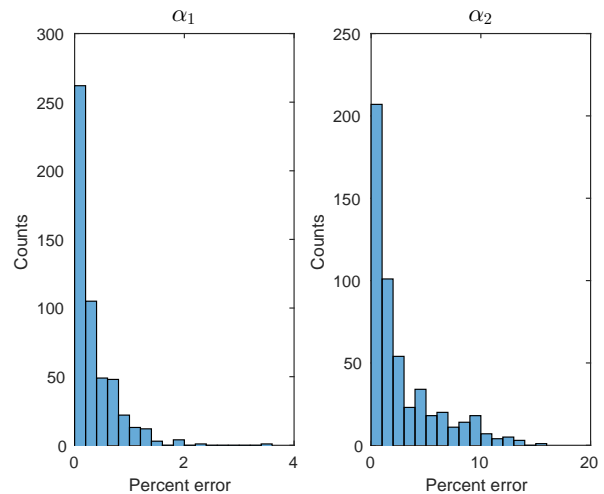}
\caption{Error histograms for the $\alpha$ coefficients of two basis fields, using an artificial neural network with 1 hidden layer consisting of 10 neurons. The mean errors are 0.34$\%$ and 2.74$\%$ while the median errors are 0.20$\%$ and 1.29$\%$ for $\alpha_1$ (ellipsoidal blob) and $\alpha_2$ (flux rope) respectively. More parameters can be found in table \ref{table:params}.}
\label{fig:twoalphas_reg}
\end{figure}

In terms of equation (\ref{eqn:parameters}), we assign $\alpha_1$ to $B_0$ of the magnetic ellipsoidal blob and $\alpha_2$ to $B_0$ of the magnetic flux rope. $\alpha_1$ was varied from 5 to 6 T (defocusing) in steps of 0.01 T while $\alpha_2$ was varied from 2.01 to 3 T in steps of 0.03 T, and all other parameters were kept constant. Radiographs of 50 by 50 pixels were generated for each configuration. As mentioned earlier, 70$\%$ of these radiographs were randomly chosen to train the artificial neural network, 15$\%$ were randomly assigned to the validation set to prevent overfitting, and the trained artificial neural network was used to predict the $\alpha_1$ and $\alpha_2$ values on the remainder $15\%$ of the radiographs. The errors, defined as $\left|\frac{\mathrm{predicted\:value}-\mathrm{actual\:value}}{\mathrm{actual\:value}}\right|$, are plotted in Fig. \ref{fig:twoalphas_reg}. We see that nearly all the errors are less than $5\%$, suggesting that the full scale implementation outlined in section \ref{subsec:recon} will work given enough basis fields. Though there are some undesirable outliers in $\alpha_2$, it is likely to be a result of inadequate data rather than a flaw in the artificial neural network method. This will be discussed in section \ref{subsec:data_accuracy}.

\subsection{Obtaining $\mathbf{B}$ field parameters from a magnetic ellipsoidal blob}
\label{subsec:blobparams}

\begin{figure}
\centering
\subfloat[ ]{\includegraphics[width=0.26\textwidth]{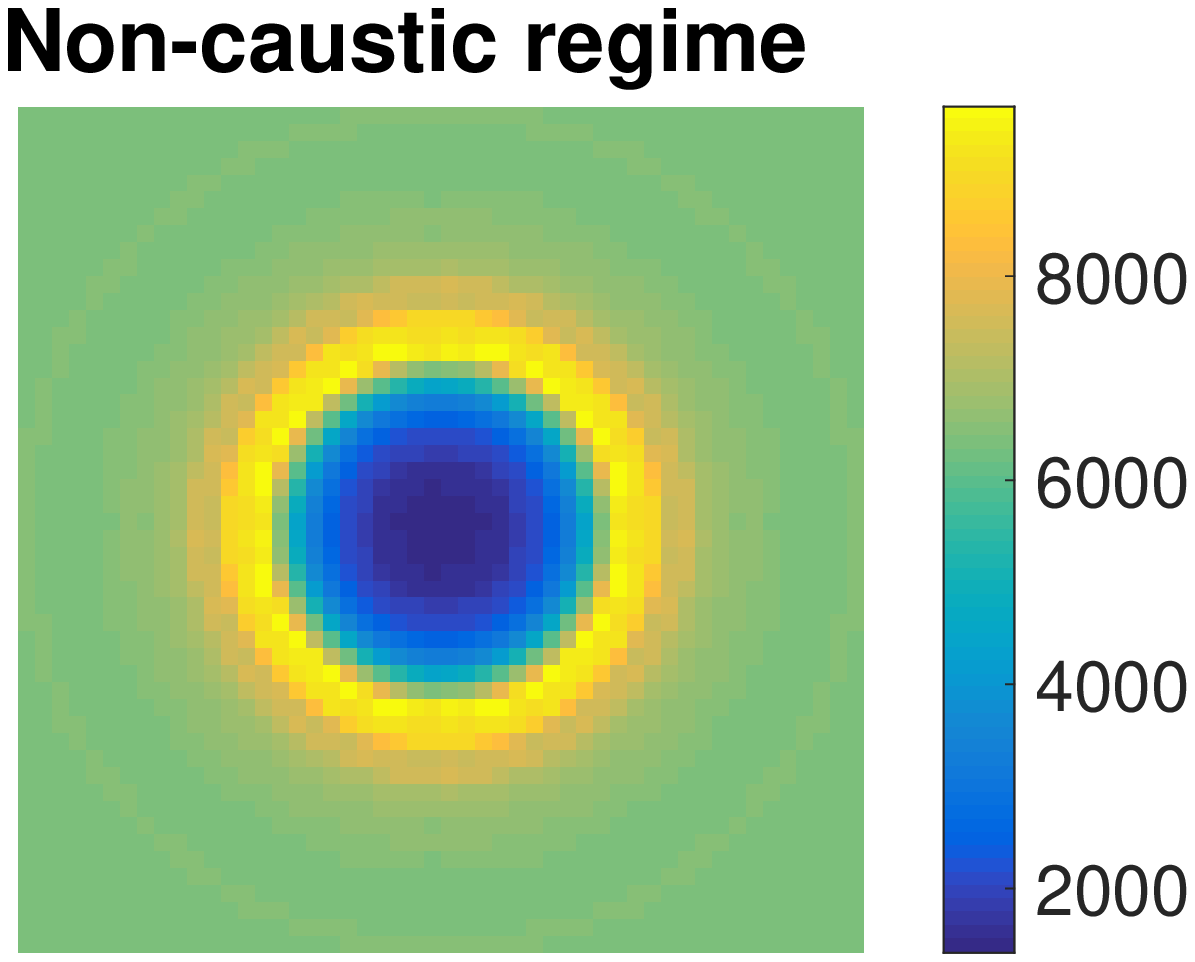}}
\subfloat[ ]{\includegraphics[width=0.26\textwidth]{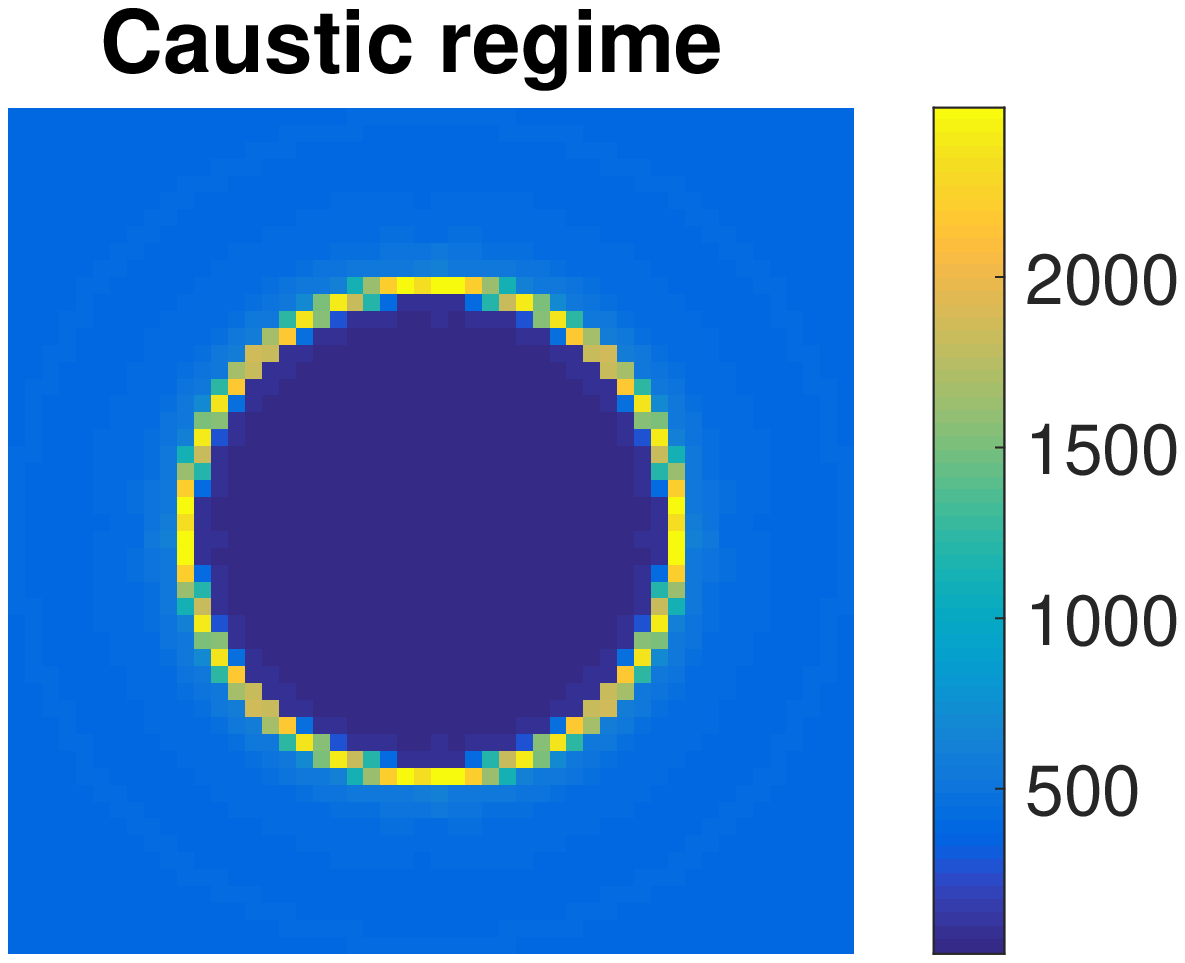}}
  \caption{(a) Radiograph for a magnetic ellipsoidal blob at B = 0.1 T, $\sigma$ = 1. This is in the non-caustic regime, where the ring around the center is smeared out. (b) Radiograph for a magnetic ellipsoidal blob at B = 0.3 T, $\sigma$ = 1. This is in the caustic regime, where most of the protons fall into a very thin ring. The scales are in arbitrary units.}
\label{fig:caustic}
\end{figure}

\begin{figure}
\includegraphics[width=0.5\textwidth]{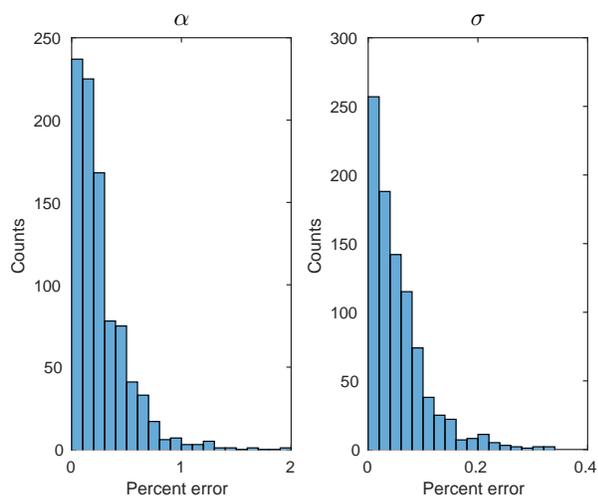}
\caption{Error histograms for $\alpha$ and $\sigma$ for a magnetic ellipsoidal blob, using an artificial neural network with 1 hidden layer consisting of 50 neurons. The mean errors are 0.26$\%$ and 0.05$\%$ while the median errors are 0.20$\%$ and 0.04$\%$ for $\alpha$ and $\sigma$ respectively.  More parameters can be found in table \ref{table:params}.}
\label{fig:alphasigma_reg}
\end{figure}

In this subsection, we demonstrate that (i) the artificial neural network method works in the non-linear regime, and (ii) the parameter retrieval concept in section \ref{subsec:retrieval} can be done. Here, we retrieve the field strength coefficient $\alpha$ and the scaling factor $\sigma$.

Radiographs for a magnetic ellipsoidal blob were generated with $\alpha$ (representing $B_0$ in equation (\ref{eqn:blob})) ranging from 0.1 to 0.3 T (defocusing) in steps of $2\times 10^{-4}$ T and $\sigma$ ranging from 0.9 to 1 in steps of 0.02. This spectrum of $\alpha$ spans both the caustic and non-caustic regime, as can be seen by the radiographs plotted in Fig. \ref{fig:caustic}. The histogram of errors are plotted in Fig. \ref{fig:alphasigma_reg}. We can see that the average errors are well below $1\%$, suggesting that artificial neural networks can be used for parameter retrieval, an alternative to reconstructing entire magnetic fields. We also see that this method works in the non-linear regime, where $\mu\approx 2$.

\begin{figure}
\includegraphics[width=0.5\textwidth]{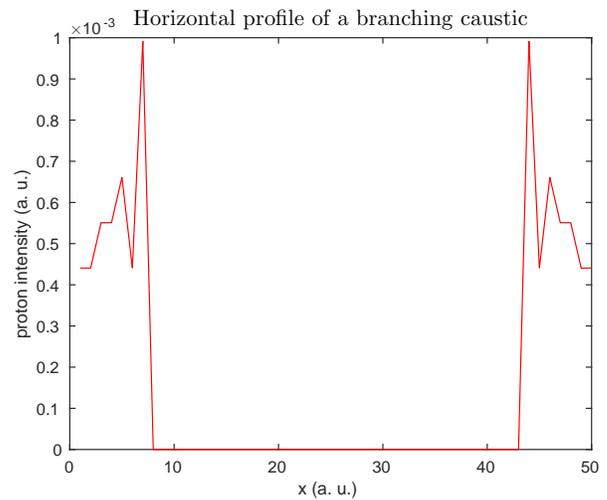}
\caption{Horizontal profile of a radiograph for a magnetic ellipsoidal blob at 1.5 T, the branching caustics regime. Notice that there are two maxima in the intensity profile, instead of one in the case of the caustic regime.}
\label{fig:branchingprofile}
\end{figure}

\begin{figure}
\includegraphics[width=0.5\textwidth]{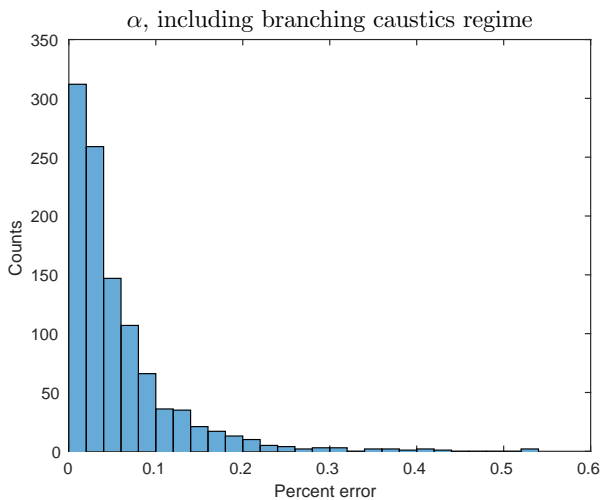}
\caption{Error histogram of $\alpha$ for a magnetic ellipsoidal blob spanning the linear, caustic and branching caustic regime, using an artificial neural network with 1 hidden layer consisting of 10 neurons. The mean error is 0.06$\%$ while the median error is 0.04$\%$.  More parameters can be found in table \ref{table:params}.}
\label{fig:branching}
\end{figure}

\subsection{Branching caustics}
\label{subsec:branching_caustics}

The power diagram method \cite{kasim} gives relative errors of more than $10\%$ in the branching caustics regime. Here, we show that the artificial neural network method is flexible enough to accommodate this scenario. We extend the range of field strengths in section \ref{subsec:blobparams} to range from 0.1 T to 1.5 T in steps of $2\times 10^{-4}$ T, spanning the linear, caustic and branching caustic regime. As an illustration, the horizontal profile of the radiograph at 1.5 T (branching caustics regime) is plotted in Fig. \ref{fig:branchingprofile}. The error histogram is plotted in Fig. \ref{fig:branching} and we can see that all the errors are well below $1\%$.

\subsection{Effect of noise on accuracy}

\begin{figure*}
\makebox[\textwidth][c]{\includegraphics[width=1.25\textwidth]{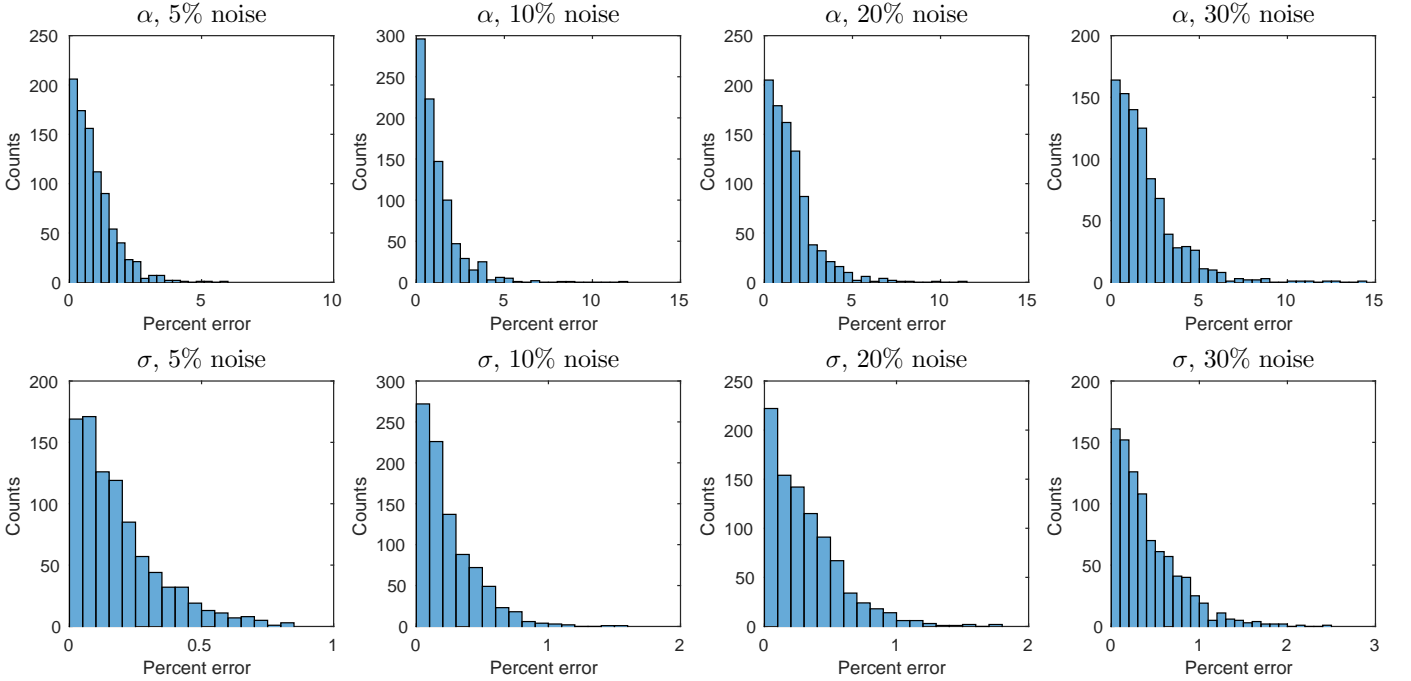}}
\caption{Error histograms of $\alpha$ and $\sigma$ when 5, 10, 20, and 30 percent noise is introduced into the radiographs for a magnetic ellipsoidal blob. The artificial neural network configurations are: 8 hidden layers with 80 neurons per layer, 5 hidden layers with 50 neurons per layer, 6 hidden layers with 70 neurons per layer and 7 hidden layers with 100 neurons per layer for 5, 10, 20, and 30 percent noise respectively. For $\alpha$, the mean errors are 0.92$\%$, 1.14$\%$, 1.49$\%$ and 1.91$\%$ while the median errors are 0.73$\%$, 0.82$\%$, 1.19$\%$ and 1.46$\%$ for 5, 10, 20, and 30 percent noise respectively. For $\sigma$, the mean errors are 0.18$\%$, 0.24$\%$, 0.31$\%$ and 0.41$\%$ while the median errors are 0.14$\%$, 0.18$\%$, 0.25$\%$ and 0.31$\%$ for 5, 10, 20, and 30 percent noise respectively. Notice that it takes an increase from 5$\%$ noise to 30$\%$ noise in order to roughly double the mean and median errors, indicating that the artificial neural network method is robust to noise.  More parameters can be found in table \ref{table:params}.}
\label{fig:alphasigma_noise}
\end{figure*}

So far we have shown that artificial neural networks trained on noise-free radiographs can retrieve quantities from noise-free radiographs with a high accuracy. We proceed to explore the changes in accuracy when noise is introduced into all the radiographs (training, validation and testing sets). Suppose a pixel in the radiograph has a value of $\chi$ and we want to introduce random noise of $x\%$. Then each pixel is replaced by a random value from a Gaussian distribution with a mean of $\chi$ and a standard deviation of $\chi\times x\%$. This was done for $x$ = 5, 10, 20 and 30 percent on the radiographs in section \ref{subsec:blobparams}, and the entire process of training and prediction was repeated. The error histograms for $\alpha$ and $\sigma$ are plotted in Fig. \ref{fig:alphasigma_noise}. We notice that for both quantities, it takes an increase from 5$\%$ to 30$\%$ noise in order to roughly double the mean and median errors. This demonstrates the robustness of artificial neural networks to noise, although noise does occasionally cause very high errors. It is worth noting that the right model to use is Poisson noise, but that model approximates Gaussian noise for a large number of particles per pixel, which is true in our case. The relationship between errors and input noise for various configurations of artificial neural networks is further explored in \cite{nwnoise1,nwnoise2,nwnoise3,sizenoise}.

\subsection{Effect of the amount of training data on accuracy}
\label{subsec:data_accuracy}

\begin{figure}
\includegraphics[width=0.5\textwidth]{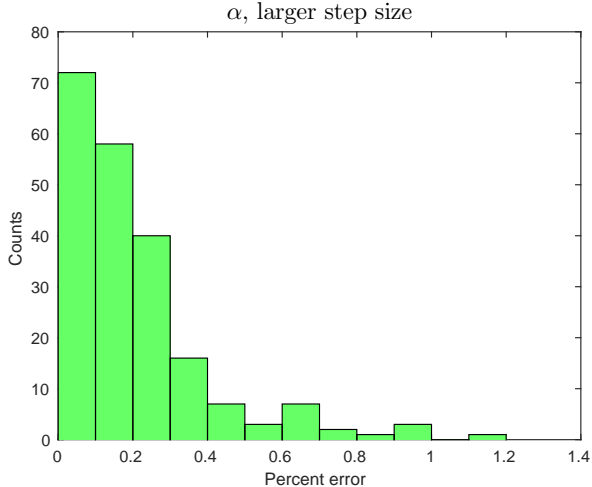}
\caption{Error histogram in $\alpha$ for a magnetic ellipsoidal blob (more parameters in table \ref{table:params}) when the step size is increased by a factor of 5, leading to less data. The mean error is 0.20$\%$ and the median error is 0.16$\%$, using an artificial neural network with 7 hidden layers with 40 neurons per layer.}
\label{fig:stepsize}
\end{figure}

While the artificial neural network method seems promising so far, it is reliant on the large amounts of training data (specifically, the amount of information in the data, or information entropy) for its accuracy. To elucidate this fact, we generated data for a magnetic ellipsoidal blob and varied only $\alpha$ between the values 0.1 to 1.5 T, similar to the scenario in section \ref{subsec:branching_caustics}, except with a larger step size of $10^{-3}$ T. The error histogram is plotted in Fig. \ref{fig:stepsize}. In comparison to Fig. \ref{fig:branching}, we see that having a larger step size and thus having less information leads to an increase in errors. This, combined with the universality of the artificial neural network proved in \cite{hornik}, suggests that extreme outliers in errors can be overcome by generating more data that increases the information entropy of the data set and re-training the neural network. The relationship between errors and size of data set for various configurations of artificial neural networks is further explored in \cite{nwsize1,nwsize2,nwsize3,sizenoise}.

\subsection{Limitations of proton radiography and the need for proton tomography}

Proton radiographs do not necessarily form one-to-one relationships with field structures: Suppose that at the edge of a plasma that is facing the proton beam, there is a very strong $\mathbf{B}$ field that deflects the incoming protons before these protons could penetrate further. Then the radiograph formed is independent of the $\mathbf{B}$ fields in the remainder of the plasma, because no protons interact with it. Due to the lack of information in the radiographs, no method can fully reconstruct the $\mathbf{B}$ fields. As such, there is a need to modify the experimental set-up to capture more information from the $\mathbf{B}$ field.

\begin{figure}
\makebox[\columnwidth][c]{\includegraphics[width=1.15\columnwidth]{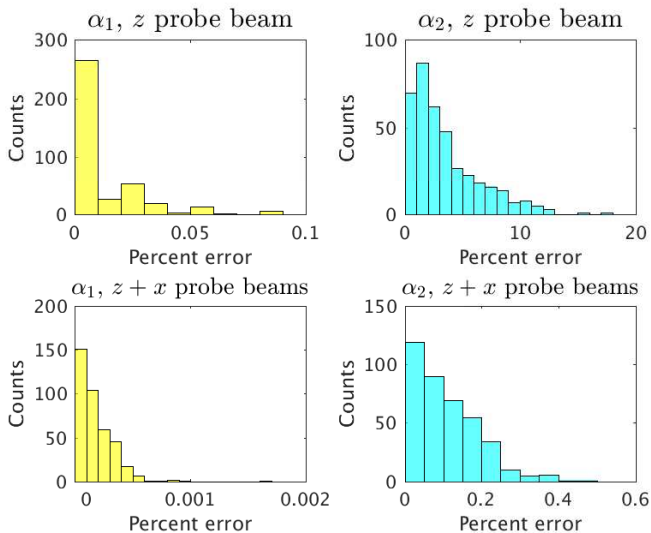}}
\caption{Error histograms of $\alpha_1$ and $\alpha_2$ for two scenarios. Top panel: Using radiographs generated by proton beams in the $z$ direction as training data, the mean errors are 1.20$\times 10^{-2}\%$ and $3.48\%$ while the median errors are 3.59$\times 10^{-3}\%$ and $2.50\%$ for $\alpha_1$ and $\alpha_2$ respectively, using an artificial neural network with 1 hidden layer consisting of 30 neurons. (more parameters in table \ref{table:params}). The errors for $\alpha_2$ (field strength of the flux rope) are high because the field structure associated with $\alpha_1$ (ellipsoidal blob) is deflecting many protons away from the flux rope, causing a lack of information in the resulting radiographs. Bottom panel: Using radiographs generated by proton beams in the $z$ and $x$ directions as training data, the mean errors are $1.81\times 10^{-4}\%$ and $0.11\%$ while the median errors are 1.41$\times 10^{-4}\%$ and $8.68\times 10^{-2}\%$ for $\alpha_1$ and $\alpha_2$ respectively, using an artificial neural network with 1 hidden layer consisting of 110 neurons. We see that upon including data from the proton beam in the $x$ direction, more information for both field structures is added to the data set and errors reduce by at least an order of magnitude.}
\label{fig:tomography}
\end{figure}

One possible way to capture more information is to include more probe beams in different directions (tomography). As an example, consider two adjacent field structures, the ellipsoidal blob (field strength parameter assigned to $\alpha_1$, ranging from 9 to 9.25 T in steps of 0.005 T) and the flux rope (field strength parameter assigned to $\alpha_2$, ranging from 0.3 to 0.4 T in steps of 0.002 T), with the former obscuring the latter in the $z$ direction by 0.5 mm. When the artificial neural network method was used on radiographs due to a beam fired in the $z$ direction, the errors in the retrieved field strength of the flux rope are very high (top panel, Fig. \ref{fig:tomography}) due to the lack of protons probing the field structure. When another probe beam was fired in the $x$ direction (with $x$ velocity of $10^6$ ms$^{-1}$ and $y,\:z$ velocities ranging from -2$\times 10^5$ ms$^{-1}$ to 2$\times 10^5$ ms$^{-1}$) and the radiographs were used in addition to the ones from the probe beam in the $z$ direction, errors for both field strengths decrease by at least an order of magnitude (bottom panel, Fig. \ref{fig:tomography}). This demonstrates two facts: (i) the artificial neural network method (and any other method) cannot fully reconstruct magnetic fields if the radiographs carry insufficient information; (ii) including more information decreases errors, even if the field structure is not obscured, as can be seen by the reduction in errors for $\alpha_1$ in Fig. \ref{fig:tomography}. We hope this will inspire future work on theoretical error bounds in artificial neural networks given the lack of information in the data set.

\subsection{Comparison with the existing radiograph inversion techniques}

The artificial neural network method addresses the two gaps in existing reconstruction techniques, by being able to work in the non-linear regime (such as the branching caustic regime), and being able to produce 3-D reconstruction of the magnetic field. While existing inversion techniques rely on the paraxial limit for simplicity, the artificial neural network technique does not rely on such a limit, and in fact would benefit more if the paraxial limit was not used--The protons should ideally have non-zero velocities in the $x$ and $y$ directions so that it will be deflected by $\mathbf{B_\textit{z}}$, allowing the artificial neural network to capture more information and thus reconstruct the magnetic fields more accurately. 

Also, existing techniques assume that the protons move in a straight line within the plasma, but this assumption does not hold when the $\mathbf{B}$ field is so strong that deflection occurs within the plasma. As a result, the existing techniques will inevitably fail in the limit of extremely large $\mathbf{B}$ fields. In comparison, the neural network method will work because it does not require this assumption. 

One major benefit of using artificial neural networks is the long-run computational cost savings. Generating each set of 50 by 50 pixel radiographs (one radiograph for each variation of parameters) takes on the order of hours/days using 16 cores on one node of the Arcus Phase B supercomputer \cite{ARC}. Training the artificial neural network takes on the order of minutes/hours when using a single GPU on the Arcus Phase B, for neural networks with up to 10 hidden layers, with each layer consisting of up to 150 neurons. Reconstruction takes on the order of seconds without using any parallel processing/GPU. If this project were to go full-scale, we can see that most of the computational cost is in the generation of training data and the training of the artificial neural network, which is a one-off cost. In comparison, existing methods of reconstruction have a recurring cost. As such, over the long run, if the artificial neural network is used to invert sufficiently many radiographs, the artificial neural network method is computationally more efficient.

However, the artificial neural network method has some drawbacks. For example, the overall accuracy of the artificial neural network can only be determined empirically, whereas error-propagation can be performed for existing techniques. While the artificial neural network method will allow for computational cost savings over the long run, the minimal start-up computational cost to get it working for a non-trivial field structure is quite high, because the artificial neural network must be trained with many basis fields before it can be used. This is in contrast to existing techniques, where any field, as long as the assumptions are met, can be imaged with the computational cost of solving a differential equation.

\section{Conclusions and future work}
\label{sec:future}

In conclusion, we have reviewed existing techniques on analyzing $\mathbf{B}$ fields from proton radiography, and the basics of artificial neural networks. Using the fact that artificial neural networks are highly flexible function approximators, we proposed for the first time the idea of using artificial neural networks to reconstruct arbitrary $\mathbf{B}$ fields and retrieve important field parameters.

Via simulations, we showed that an artificial neural network can reconstruct $\mathbf{B}$ fields that can be expressed as linear combinations of two fields, and retrieve useful quantities of $\mathbf{B}$ fields such as characteristic lengths. We also explored the effects of noise and size of data set on the accuracy of the artificial neural network, and found that artificial neural networks are robust to noise. Artificial neural networks can accommodate a wide variety of scenarios and assumptions which existing techniques cannot, such as the branching caustics part of the non-linear regime. We also highlighted the need for proton tomography as certain field structures cannot be reconstructed fully due to the lack of information from a single radiograph.

As the usage of artificial neural networks in diagnosing $\mathbf{B}$ fields in high energy density plasmas is new, there are many avenues where this work can be developed further. There are at least three ways to improve the accuracy of the artificial neural network: (i) experiment with other types of artificial neural network architecture. For example, convolutional neural networks are a type of feedforward artificial neural network where the connections between neurons are inspired by the animal visual cortex \cite{convnet} and as such, perform very well in image recognition. Since radiograph inversion involves image recognition, convolutional neural networks could offer better performance than the fully connected (dense) feedforward neural network used in this paper. Recurrent neural networks are a class of artificial neural networks where the neuron connections form directed cycles, and such architecture has advantages in analyzing time series data. We could use recurrent neural networks on a time series of proton radiographs to shed light on the dynamics of the $\mathbf{B}$ field and hence the plasma. (ii) Include energy-resolved radiographs. In our simulations, we only looked at the spatial distribution of the protons, so including extra information on the proton energies could improve accuracy. (iii) Study the effects of the number of pixels on accuracy.  It is interesting to note that promising results were obtained despite the low resolution of the radiographs (50 by 50 pixels). Understanding the effects of discretization noise could help us determine the quality of radiographs to be generated in order to train an artificial neural network to a specific accuracy.

Furthermore, the artificial neural network approach can be extended to similar systems, such as diagnosing electric fields in plasmas or characterizing micromagnetic patterns in magnetic media via electron scattering \cite{micromag}. Finally, a full scale implementation of an artificial neural network that can reconstruct any $\mathbf{B}$ field is a possibility we can look forward to.

\begin{acknowledgements}

The authors would like to acknowledge the support from the plasma physics HEC Consortium EPSRC grant number EP/L000237/1, as well as the Hartree Centre, Daresbury Laboratory, Central Laser Facility,  and the Scientific Computing Department at the Rutherford Appleton Laboratory. The authors would like to acknowledge the use of the University of Oxford Advanced Research Computing (ARC) facility in carrying out this work \cite{ARC}. N.C. acknowledges financial support from the Singapore government. M.F.K. gratefully thanks the Indonesian Endowment Fund for Education for its support. M.C.L. thanks the Royal Society Newton International Fellowship for support. The authors acknowledge support from OxCHEDS and P.A.N. for his William Penney Fellowship with AWE plc. 

\end{acknowledgements}

\appendix

\bibliography{biblio}

\end{document}